# Combining noisy well data and expert knowledge in a Bayesian calibration of a flow model under uncertainties: An application to solute transport in the Ticino basin

Authors:
Emily A. Baker[a,b], Sauro Manenti[a], Alessandro Reali[a,c],
Giancarlo Sangalli[b,c], Lorenzo Tamellini[c], Sara Todeschini[a]

[a]Department of Civil Engineering and Architecture (DICAr) and Research Centre on Water (CRA), University of Pavia, Via Ferrata, 3, 27100, Pavia, Italy

[b]Department of Mathematics, University of Pavia, Via Ferrata, 5, 27100, Pavia, Italy

[c]Institute of Applied Mathematics and Information Technologies "E. Magenes", National Research Council (CNR-IMATI), Via Ferrata 5/A, 27100, Pavia, Italy

**Corresponding Author:** Sauro Manenti

Emily A. Baker: https://orcid.org/0000-0003-3443-5419, emily.baker@unipv.it
Sauro Manenti: https://orcid.org/0000-0002-5467-9640, sauro.manenti@unipv.it
Alessandro Reali: https://orcid.org/0000-0002-0639-7067, alereali@unipv.it
Lorenzo Tamellini: https://orcid.org/0000-0002-8008-0359, tamellini@imati.cnr.it
Giancarlo Sangalli: https://orcid.org/0000-0002-5642-1969, giancarlo.sangalli@unipv.it
Sara Todeschini: https://orcid.org/0000-0002-5498-4370, sara.todeschini@unipv.it

**Acknowledgements:**
Support for this research was provided by a grant through Regione Lombardia, POR FESR 2014-2020 - Call HUB Ricerca e Innovazione, Progetto 1139857 CE4WE: Approvvigionamento energetico e gestione della risorsa idrica nell'ottica dell'Economia Circolare (Circular Economy for Water and Energy). A. Reali and S. Manenti were also partially supported by the Italian Ministry of Education, University and Research (MIUR) through the PRIN project XFAST-SIMS (No. 20173C478N). L. Tamellini was supported by the PRIN 2017 project 201752HKH8 "Numerical Analysis for Full and Reduced Order Methods for the efficient and accurate solution of complex systems governed by Partial Differential Equations (NA-FROM-PDEs)". The authors are grateful to Regione Lombardia and the Agenzia Regionale per la Protezione dell'Ambiente (ARPA) for access to monitoring data. In addition, the authors wish to thank Professor Giorgio Pilla (University of Pavia) for spring/fontanili data, aquifer base data, and for help with the aquifer conceptualization, Dr. Maurizio Gorla (Gruppo CAP Milano) and Fasani Maurizio (Studio Trilobite, Via San L. Beccari n. 2 - Gropello Cairoli) for stratigraphy and borehole log data.

## Abstract

Groundwater flow modeling is commonly used to calculate groundwater heads, estimate groundwater flow paths and travel times, and provide insights into solute transport processes within an aquifer (e.g. Ntona et al., 2022, You et al., 2020). However, the values of input parameters that drive groundwater flow models are often highly uncertain due to subsurface heterogeneity and geologic complexity in combination with lack of measurements/unreliable measurements. This uncertainty affects the accuracy and reliability of model outputs. Therefore, parameters' uncertainty must be quantified before adopting the model as an engineering tool. In this study, we model the uncertain parameters as random variables and use a Bayesian inversion approach to obtain a posterior, data-informed, probability density function (pdf) for them: in particular, the likelihood function we consider takes into account both well measurements and our prior knowledge about the extent of the springs in the domain under study. To keep the modelistic and computational complexities under control, we assume Gaussianity of the posterior pdf of the parameters. To corroborate this assumption, we run an identifiability analysis of the model: we apply the inversion procedure to several sets of synthetic data polluted by increasing levels of noise, and we determine at which levels of noise we can effectively recover the "true value" of the parameters. We then move to real well data (coming from the Ticino River basin, in northern Italy, and spanning a month in summer 2014), and use the posterior pdf of the parameters as a starting point to perform an Uncertainty Quantification analysis on groundwater travel-time distributions.

## 1. Introduction

Groundwater mathematical models are often used to simulate groundwater heads and flows, estimate groundwater travel time, and increase understanding of solute transport processes in an aquifer system (e.g. Ntona et al., 2022, You et al., 2020). Unfortunately, such models are often associated with large uncertainties in model inputs (such as hydraulic conductivity, porosity, and recharge rate) and in other parameters used to configure the boundary conditions (e.g. Bianchi Janetti et al., 2019). These uncertainties can arise from a combination of factors, including the complex and heterogeneous nature of the aquifer system, as well as insufficient measurements of aquifer properties, groundwater heads, and boundary conditions. Such parameter uncertainties can make it difficult to accurately estimate travel time and solute transport within an aquifer. To overcome such challenges, a common approach is to model uncertain parameters as random variables and consider an Uncertainty Quantification (UQ) approach to the prediction problem, divided in two steps. In the first one, we employ Bayesian inversion (Stuart, 2010) to reduce the uncertainty on the parameters by incorporating the data at hand, or more precisely, by computing the posterior probability density function (pdf) of the uncertain parameters. In the second step, we perform a forward UQ analysis (Ghanem et al., 2017), using a relatively straight-forward Monte Carlo method: we generate several random values of the parameters (according to their posterior pdf), and compute for each set of values first the groundwater flow and then the corresponding travel times of solute particles released in the domain; finally, we perform statistical analyses on the sets of travel times thus obtained, to provide robust estimates of such times.

The Bayesian approach we employ in the first step introduces a non-standard aspect in that our likelihood function incorporates not only raw data (specifically, groundwater well measurements) but also expert knowledge on the amount of land surface that is predicted to be covered by springs (ideally around 1% but in any case not exceeding the range 0-2%, based on previous "qualitative" knowledge about the peculiarities of the flow in study area). To simplify the subsequent forward UQ analysis, we introduce the widely adopted assumption that the posterior pdf is approximately Gaussian (Bui-Thanh et al., 2013; Piazzola et al., 2021). This assumption is well-suited if the posterior pdf is unimodal, symmetric, and well-peaked, or equivalently if the log-posterior function has a unique, narrow minimum at the center of ellipsoidal isolines. To assess the validity of this assumption, we carry out an identifiability analysis (Piazzola et al, 2021; Guillaume et al., 2019; Raue et al., 2009), where we repeatedly perform the Bayesian inversion using different sets of synthetic data, i.e., artificial groundwater head data that were generated by first running the groundwater flow model using a known set of input parameter values and then polluting the results by adding Gaussian noise to mimic measurement errors. The goal of these preliminary tests is to verify whether the simplified Bayesian procedure (with the Gaussian assumption) reasonably identifies the true values of the uncertain parameters (i.e., if the peak of the posterior pdf is close to the true value, and its standard deviation is not too large). The different sets of data we consider are generated by adding Gaussian noise with increasingly large standard deviation, to assess up to which level of noise the procedure is reliable, before running it for the observed data.

The groundwater flow model to which this procedure is applied in this study is in the Ticino groundwater basin as shown in Figure 1, which underlies an important agricultural region in Northern Italy (Baker et al., 2022). Like many groundwater models, there is uncertainty surrounding the hydraulic conductivity of the aquifer and uncertainty due to the model boundary conditions (e.g., how the rivers, springs, and model edges are constructed). In addition, there is also uncertainty associated with the amount of aquifer recharge that occurs due to the extensive irrigation activities in the region. While the flood irrigation techniques that are applied to the rice fields are known to contribute recharge to the superficial aquifer system, the magnitude of this contribution is highly uncertain. Lastly, there is additional uncertainty introduced due to the limitations of the observed groundwater head data. While there are twenty-two groundwater wells located in the study area, their heads were not sampled at the same time. Rather, the heads were sampled over the span of approximately one month during August and September of 2014. Furthermore, some levels may not have completely recovered post-pumping prior to measurement, adding additional uncertainty. Therefore, by applying the above method, this study aims to overcome this combination of uncertainties and estimate the accuracy to which the groundwater model can predict travel times.

We close this introduction by mentioning that analyses of travel times of passive solute transport have been proposed in several works in the UQ literature. The main difficulty of these analyses arises from the fact that passive transport is typically described by hyperbolic PDEs (e.g., linear conservation laws or more complicated versions), which are known to be hard to solve efficiently in a UQ context (resorting, e.g., to surrogate modeling techniques) due to the fact that the solution of the PDE (in this case, the concentration of the solute at each point in space and time) does not depend smoothly on its uncertain parameters (typically, the transport field). As a consequence, Monte Carlo methods such as the one briefly sketched above, which are insensitive to this problem, are often employed (Salandin et al., 1998; Riva et al., 2006; Charrier, 2015). To reduce the cost of Monte Carlo methods, several strategies can be conceived. Müller et al. (2011) propose approximating the Darcy velocity field by surrogate modeling and then applying MC only to the transport part of the problem. Other works (Müller et al., 2014; Tesei, 2016; Crevillén-García et al., 2017) consider variance reduction by Multi-Level Monte

Carlo. A different approach, based on Markovian Velocity Processes is proposed in Meyer et al. (2013). Liao et al. (2016) propose to rewrite the problem in a more convenient form and then apply suitable surrogate modeling techniques. In our work, the cost of the Monte Carlo method is mitigated by the fact that the uncertainty on the parameters left to explore after the Bayesian inversion process is reduced by a factor at least 10 with respect to the prior information around 90% of the times. The novelty of this work thus does not reside in the Monte Carlo method per se, but in the "non-standard" formulation of the Bayesian inversion to incorporate expert knowledge, combined with the application to a real test case.

The rest of this paper is organized as follows: Section 2.1 describes the geography of the study area; Section 2.2 describes the numerical methods used to compute the fluid flow and the trajectories of the solute particles (using MODFLOW and MODPATH, respectively); Section 2.3 provides details on the model used (boundary conditions, initial conditions, geometry, computational grid, etc.). The Uncertainty Quantification method is described in Section 3: more specifically, Sections 3.1 through 3.3 deal with the Bayesian inversion approach (computation of the nominal value of the parameters and of their covariance matrix); Section 3.4 presents the identifiability analysis to assess the validity of the Bayesian inversion procedure; Section 3.5 describes the Forward UQ analysis for the travel times computed by MODPATH. Computational results are then reported in Section 4: Section 4.1 discusses the results of the identifiability analysis; Section 4.2 applies the Bayesian inversion to the real data; the forward UQ analysis for travel times is performed on synthetic datasets in Section 4.3, while real data are considered in Section 4.4. Finally, conclusions are drawn in Section 5.

## 2. Groundwater and particle flow model description

**2.1 Study Area**

The study area is in the Po Plain of Northern Italy and encompasses the southern half of the Ticino basin as shown in Figure 1a-b. The study area is approximately 501.5 km$^2$ and consists of the portion of the Ticino basin between the town of Abbiategrasso in the north to the city of Pavia in the south. The Ticino River flows south through the basin and joins the westward flowing Po River at the southern end of the study area. The length of the Ticino River within the model domain is 55.7 km, while the length of the Po River along the southern model edge is 9.1 km. Natural and human enhanced springs commonly occur in the basin (Regione Lombardia, 2007; Regione Lombardia, 2013b; De Luca et al., 2014; Balestrini et al., 2021), especially where an abrupt change in land surface elevation (20-30 m) occurs between the higher elevation plains along the edges of the basin to the lower elevation river valley in the basin center. The basin also contains a network of typically unlined canals and irrigation ditches that provide water sourced from the upstream part of the river and springs for agricultural purposes, dominantly for rice field irrigation, which constitutes 29% of land use in the area (Regione Lombardia, 2019). These irrigation activities contribute to the aquifer recharge in the region, with 40-50% or more of the irrigation water recharging the underlying superficial aquifer (Regione Lombardia, 2008; Lasagna et al., 2020). In addition to recharge from canal leakage and flood irrigation in rice fields, the superficial aquifer is also recharged by precipitation. Within the study area, the superficial unconfined aquifer ranges in thickness from about 35 - 109 m and consists of proximal braid plain deposits. These deposits are from the middle-late Pleistocene and consist of gravel within a sandy matrix (De Caro et al., 2020). Previous studies in region indicate the hydraulic conductivity of the superficial aquifer ranges from about $2\times10^{-4} - 1\times10^{-3}$ m/s, but can vary spatially, with values as low as 1×10$^{-5}$ – 1×10$^{-4}$ m/s and as high as 2×10$^{-3}$ – 5×10$^{-3}$ m/s (De Caro et al., 2020; Lasagna et al., 2020). Underlying aquifer units in the study area are separated from the superficial aquifer by confining layers, but these units are not considered in the present study.

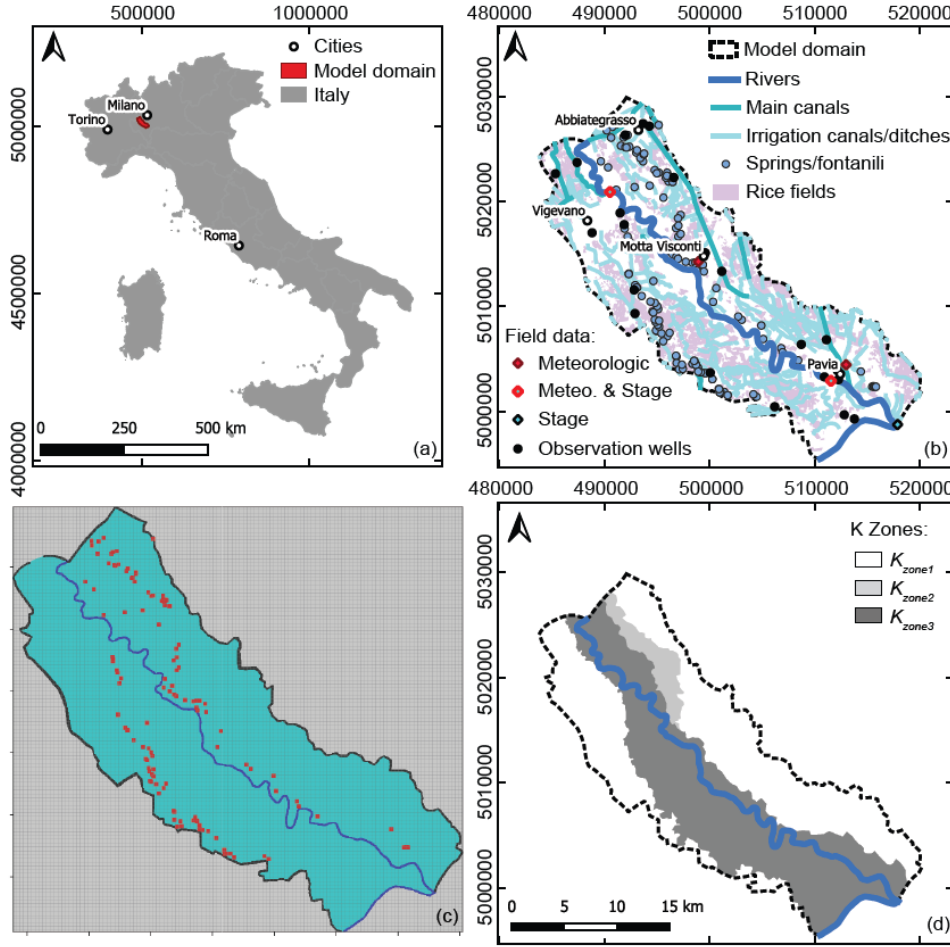

**Figure 1**. (a) Location of the model domain in Northern Italy. (b) Map of main hydrologic features (rivers, springs, canals), observation wells, monitoring stations, and rice fields within the model domain. (c) Model grid depicting the boundary conditions applied to the grid cells to represent the hydrologic features. Drain cells representing springs/fontanili are red, constant head cells representing the rivers are blue, general head boundary cells at the edges of the active model domain are black, and cells receiving recharge (from precipitation, irrigation and/or canal leakage) are teal. Rice field locations (additional drain boundaries) are not shown but can be seen in 1b. (d) Map depicting the extent of the three hydraulic conductivity zones used in the model.

## 2.2 Numerical Model

The groundwater model used in this study is MODFLOW 6 (Langevin et al., 2017; Langevin et al., 2021), an open-source code developed by the United States Geological Survey (USGS). MODFLOW simulates groundwater heads and fluxes through space and/or time by solving the governing differential equation in every model grid cell using the finite difference method (Harbaugh, 2005; Langevin et al., 2017; Langevin et al., 2021). We consider the steady-state version of the model, whose governing equation is:

$$\frac{\partial}{\partial x}\left(K_{xx}\frac{\partial h}{\partial x}\right) + \frac{\partial}{\partial y}\left(K_{yy}\frac{\partial h}{\partial y}\right) + \frac{\partial}{\partial z}\left(K_{zz}\frac{\partial h}{\partial z}\right) + W = 0 \qquad [1]$$

where:
- $K_{xx}$, $K_{yy}$, and $K_{zz}$ are the principal components of the hydraulic conductivity tensor (L/T) along the coordinate directions *x*, *y*, and *z* (vertical direction);
- *h* is the piezometric head (L);
- *W* is a volumetric flux per unit volume representing sources (*W*>0) and/or sinks (*W*<0) of water in the groundwater system ($T^{-1}$);

The governing equation (Eq. 1) is derived from the continuity equation for the conservation of mass and Darcy's law for saturated porous media. When the aquifer is unconfined, the position of the phreatic surface (i.e., the interface between the saturated and unsaturated zones) is not known a-priori and the problem becomes non-linear (Mehl, 2006; Painter et al., 2008). In this study, the Newton-Raphson method is used to solve the system of groundwater flow equations rather than the standard MODFLOW method (Picard iterations) because it is more stable and converges more reliably for problems where the water table traverses multiple cell layers due to factors such as complex geology and steep terrain (as in our case) (Niswonger et al., 2011; Langevin et al., 2017).

After MODFLOW is used to compute the groundwater flow field, MODPATH version 7 is then used to perform particle tracking (Pollock, 1988; Pollock, 2016). This operation is done by releasing particles at the land surface in cells specified by the user, computing the trajectories of such particles by integrating with a semi-analytic method the kinematic equation of their motions (pure advection transport) and from this deriving the amount of time required for them to exit the computational domain. This allows us to obtain the cumulative frequency distribution of the travel times of groundwater discharging from the model (Pollock, 1994; Visser et al., 2009; Pollock 1988; Pollock 2016). Typically, this discharging groundwater exits the model through rivers, springs, wells, or other discharge boundaries within the model domain. The effects of diffusion and dispersion are ignored since they would require field data, hardly achievable on such a large domain, to calibrate the model coefficients of related terms in the transport equation (Pollock, 1994; Visser et al., 2009). We use the FloPy package (Bakker et al., 2016) in Python to write both the MODFLOW and MODPATH input files and implement the appropriate boundary conditions.

## 2.3 Model Setup

The groundwater flow model used in this study encompasses the southern half of the Ticino basin. As already mentioned, it is run as a steady-state model, focused on August/September 2014 conditions due to the data available: a single value of observed groundwater head data at 22 well locations, whose measurement times are not identical (but all included in the two months mentioned). The model domain consists of 50 x 50 m grid cells in the horizontal plane, resulting in 686 rows and 727 columns, and 3 horizontal layers of grid cells, for a total of about $6.01 \times 10^5$ active grid cells. To better capture the water table position while minimizing the computational expense, the top two layers of grid cells are each a quarter of the model thickness, while the bottom layer of grid cells is half of the model thickness. A 5 m digital terrain model (Regione Lombardia, 2015) was resampled to a 50 m resolution which was then used to set the surface elevations of the grid cells. The base of the superficial aquifer was created by interpolating data of the unconfined aquifer basal elevations (Regione Lombardia, 2022). According to a hydrogeological conceptual model that was proposed in a previous work (Baker et al., 2022), the superficial aquifer consists of three zones with differing hydraulic conductivities as depicted in Figure 1d: 1) a lower hydraulic conductivity zone ($K_{zone1}$) along the model edges in the higher elevation plains, 2) a zone with intermediate hydraulic conductivity ($K_{zone2}$) in the northeastern portion of the model domain, and 3) a higher hydraulic conductivity zone ($K_{zone3}$) in the central river valley (see Table 1 top row for the corresponding ranges). The uncertain ranges for $K_{zone1}$ and $K_{zone2}$ are the same, though it is suspected that $K_{zone2}$ is likely slightly larger than $K_{zone1}$. $K_{zone2}$ and $K_{zone3}$ only occur in the top two cell layers, while $K_{zone1}$ occurs in the lowermost cell layer across the whole model domain. These three conductivities are therefore considered uncertain, and the subject of our Bayesian inversion technique detailed in the following. Note that throughout the work we will always enforce a physically based condition that the values of conductivity that we consider when running the model are always sorted, ie., $K_{zone1} \leq K_{zone2} \leq K_{zone3}$.

Multiple types of boundary conditions are implemented in the groundwater model to represent the various hydrologic features present in the study area, shown in Figure 1c. A no-flow boundary is implemented at the bottom surface of the model domain to mimic the base of the superficial aquifer where it encounters an underlying confining layer. A general head boundary is implemented along the edge of the model domain where the edge does not align with a river. Along the northern boundary of the model domain, the edge was delineated such that it is perpendicular to the groundwater head contours such that groundwater flow is parallel to the northern model edge. A contour map of groundwater heads from August/September 2014 groundwater well data was used to assign the head values of the general head boundary along the edge of the model domain (Regione Lombardia, 2014).

A specified head boundary, imposed by the MODFLOW Constant Head Designation (CHD) package, is used to represent the Ticino and Po rivers. Water levels at 35 locations (computational sections) along the length of the Ticino River were simulated during the same period using a 1D unsteady hydraulic model of the river developed with the open-source software HEC-RAS 5.0.7 (HEC-RAS, 2019). River stage data were measured at two gauging stations (near Vigevano and Pavia) along the Ticino River and at the confluence of the Ticino and Po rivers (AIPo, 2004; AIPo, 2005; AIPo, 2020). The measured stage data from the two gauging stations along the Ticino River were used to calibrate the HEC-RAS model and assess the accuracy of the simulated stage levels. The simulated water levels were then linearly interpolated between the 35 HEC-RAS model sections to estimate a stage value for every CHD grid cell of the groundwater flow model through which the Ticino River passes. The interpolated head values were then directly applied to these grid cells using the CHD package to represent the river. Additional information on the setup and calibration of the HEC-RAS hydraulic model used to calculate the river levels can be found in Baker et al. (2022) and Cappato et al. (2022). Meanwhile, the stage along the short stretch of the Po River included in the model was assumed constant, with a value equal to that measured at the confluence. These estimated stage values are the head values assigned to the river cells using the specified head boundary condition.

The recharge boundary condition, imposed by the MODFLOW Recharge (RCH) Package, is used to apply recharge to the aquifer in the groundwater flow model. Within the model domain, recharge comes from precipitation, irrigation, and leakage from irrigation canals. The recharge rate due to precipitation was estimated by subtracting the potential evapotranspiration (PET) rate and runoff from the precipitation rate, with no recharge occurring if the PET and runoff exceeded the precipitation rate. The PET rate was calculated using the FAO-56 Penman-Monteith method (Allen et al., 1998) using meteorological data collected at three stations in the study area (ARPA, 2020) using the PyETo package in Python (Richards, 2015). The runoff rate was estimated using the Soil Conservation Service (SCS) Curve Number (CN) method (Mishra & Singh, 2003) and maps of the hydrological soil class (Regione Lombardia, 2013a) and land cover type (Regione Lombardia, 2019). Weekly recharge rates were calculated for August and September of 2014 and then averaged to obtain a recharge rate for the study period. Additional recharge was then added to the model to account for recharge due to flood irrigation within rice fields and leakage from the irrigation canal network. The amount of recharge due to irrigation activities was also subject to calibration by Bayesian inversion, due to the limited availability of data on such recharge rates. As an initial rough estimate, we assume that the recharge rate can range in the interval reported in Table 1.

**Table 1**. Uncertain ranges of each input parameter value based on a-priori information about the groundwater basin, the parameter values used to create the base case simulation to which Gaussian noise was added to generate synthetic datasets, and the estimated parameter values determined through the optimization procedure using the observed well data.

|  | $K_{zone1}$ (m/s) | $K_{zone2}$ (m/s) | $K_{zone3}$ (m/s) | $R_{irrig}$ (m/s) |
|---|---|---|---|---|
| Parameter Uncertain Ranges | 5.0e-5 – 1.0e-3 | 5.0e-5 – 1.00e-3 | 1.0e-4 – 1.0e-2 | 1.0e-10 – 1.0e-6 |
| Base case (true) parameter values for synthetic data | 6.5e-5 | 4.5e-4 | 5.0e-3 | 2.5e-8 |
| Estimated parameter values for observed data | 8.13e-5 | 8.13e-5 | 6.62e-3 | 5.69e-8 |
| Estimated standard deviation of the parameter values for observed data | 3.72e-6 | 1.92e-5 | 1.60e-3 | 2.14e-8 |
| Coefficients of variation | 0.0458 | 0.2362 | 0.2417 | 0.3761 |

The drain boundary condition is used to represent the numerous springs and fontanili (human enhanced springs) in the study area and imposed by the MODFLOW Drain (DRN) Package, an approach similar to other studies (Bianchi Janetti et al., 2019). Approximately 140 springs and fontanili have been mapped in the model domain (Regione Lombardia, 2007; Regione Lombardia, 2013b; Magri, 2020; Gardini, 2021), corresponding to 132 model grid cells. The grid cells containing mapped springs and fontanili are assigned the drain boundary condition, such that groundwater in these cells is removed from the model when the water head exceeds the elevation of the land surface. The rate of discharge from the drain cells out of the model domain is equal to the height of the water above the land surface multiplied by the drain conductance, which was set to 100 m$^2$/s. Additional drain boundaries were also placed in grid cells that contain rice fields to allow for the slight ponding of irrigation waters that can occur due to the implemented flood irrigation techniques in the region and to remove any excess irrigation recharge that results in the exceedance of these irrigation depths, which would typically be removed as runoff into the irrigation canals.

A more complete description of the model structure and the data and methods used to construct the boundary conditions can be found in Baker et al. (2022), while a more complete description of MODFLOW and the numerical implementation of its boundary conditions can be found in the program documentation (Harbaugh, 2005; Hunt & Feinstein, 2012, Langevin et al., 2017; Langevin et al., 2021).

## 3. Uncertainty Quantification (UQ) analysis

In this section we describe the work plan for UQ: first the Bayesian inversion approach to "calibrate" the parameters (technically, by computing their posterior pdf), and then the forward UQ analysis to propagate the uncertainty encoded by such pdf from the parameters to the quantities of interest of the problem, namely the particle travel times.

### 3.1 Step 1: Bayesian Inversion for the uncertain inputs

Summarizing the discussion in the previous sections on the groundwater model in the current study, the uncertain input parameters include the aquifer recharge rate due to irrigation ($R_{irrig}$) and the hydraulic conductivity values in three different zones within the study area ($K_{zone1}$, $K_{zone2}$, $K_{zone3}$), that can take values in the ranges reported in Table 1. There was also some uncertainty in the stage values of the rivers ($S_{RIV}$), the conductance of the drain cells ($C_D$), and the head value of the general head boundary at the model edges ($H_{GHB}$), but prior sensitivity analysis revealed that these uncertainties were not influential on the model results (Baker et al., 2022), and so $S_{RIV}$ and $H_{GHB}$ were set to measured/interpolated values, while $C_D$ was set to 100 m$^2$/s (Musacchio et al., 2021; Baker et al., 2022). For notational convenience, we collect the uncertain parameters in a vector $\mathbf{p} = [p_1, p_2, p_3, p_4] = [K_{zone1}, K_{zone2}, K_{zone3}, R_{irrig}]$. A-priori (i.e., without any data available), we can assume that these parameters are mutually independent and can take any value in their ranges with "equal probability", i.e., their prior distribution $\rho_{prior}$ can be considered as uniform over their ranges (another sensible assumption would be to model $K_{zone1}$, $K_{zone2}$, $K_{zone3}$ as log-uniform random variables given that their ranges span multiple orders of magnitude, as well as the classic lognormal assumption, see e.g. Ricciardi et al. (2005); a thorough comparison between the possible models would have however exceeded the scope of the current work). We then employ a Bayesian inversion approach to reduce their uncertainty, i.e., to obtain a posterior, data-informed, probability density function (pdf) $\rho_{post}$ for them. This approach relies on the Bayes formula for conditional probabilities, which dictates that:

$$\rho_{post}(\mathbf{p}) = \mathcal{L}(\mathbf{p}, \mathbf{h})\rho_{prior}(\mathbf{p})\frac{1}{C} \qquad [2]$$

where
- $\mathbf{h}^* = [h^*_1, h^*_2, \ldots, h^*_{nb\,wells}]$ are the data available; in our problem, they are the well head measurements at nb$_{wells}$ = 22 locations $x_1, x_2, \ldots, x_{nb\,wells}$,
- $\mathcal{L}(\mathbf{p}, \mathbf{h}^*)$ is the likelihood function
- $C$ is a normalization constant that guarantees that $\rho_{post}$ integrates to 1.

Informally, the likelihood function $\mathcal{L}(\mathbf{p}, \mathbf{h}^*)$ is a function that quantifies the "probability" of observing the data that we measured if the uncertain parameters had value $\mathbf{p}$, and thus encodes our information on the model and on the data. In particular, under the assumption that the measured well data $h^*_i$ are equal to the heads predicted by our MODFLOW model at location $x_i$ for some unknown values of the parameters $\mathbf{p}_{true}$, $h(x_i, \mathbf{p}_{true})$, plus a Gaussian random variable with zero mean and variance $\sigma_h^2$ (which plays the role of measurement error), i.e.,

$$h_i^* = h(x_i, \mathbf{p}_{true}) + \varepsilon_i, \varepsilon_i \sim \mathcal{N}(0, \sigma_h^2), i = 1, \ldots nb_{wells},  \qquad [3]$$

the likelihood function can be written as

$$\mathcal{L}(\mathbf{p}, \mathbf{h}^*) = \prod_{i=1}^{nb_{wells}} \frac{1}{\sigma_h \sqrt{2\pi}} e^{-\frac{(h_i^* - h(x_i, \mathbf{p}))^2}{2\sigma_h^2}}. \qquad [4]$$

Note that evaluation of the likelihood function is expensive since it entails solving the flow model for the specified values of the parameters.

At this point, we follow a common approach and further introduce the approximation that $\rho_{post}$ is a Gaussian distribution (Bui-Thanh et al., 2013; Piazzola et al., 2021). This considerably simplifies the subsequent forward UQ step (see Section 3.5) and is a reasonable approximation when the posterior pdf is symmetric, unimodal, and well-peaked. To make this approximation practical, we must only do two things, i.e., compute the mean and the covariance matrix of such a Gaussian distribution, that we call $\mu_{post}$ and $\Sigma_{post}$, respectively. We devote the next two subsections to the computation of these two objects.

### 3.2 Estimating the mean of the Gaussian approximation of the posterior pdf

The mean $\mu_{post}$ can be thought as the "nominal", "most-likely" value of the parameters after the inversion procedure, and it's therefore easy to see that it should be located where the posterior pdf has its maximum (i.e. at the mode of the posterior); therefore, we should compute

$$\mu_{post} = \mathrm{argmax}_{\mathbf{p}} \rho_{post}(\mathbf{p}) = \mathrm{argmax}_{\mathbf{p}} [\mathcal{L}(\mathbf{p}, \mathbf{h}^*) \rho_{prior}(\mathbf{p}) \frac{1}{C}] = \mathrm{argmax}_{\mathbf{p}} \mathcal{L}(\mathbf{p}, \mathbf{h}^*) \qquad [5]$$

where the last equality is true since $C$ and $\rho_{prior}$ are constants (the former by definition, the latter by our assumption that the prior is uniform) and therefore do not impact the optimization procedure. Numerically, it is more convenient to further manipulate Eq.5 by taking the negative logarithm of the likelihood and computing the mean of the Gaussian posterior as:

$$\mu_{post} = \mathrm{argmin}_{\mathbf{p}} [-\log(\mathcal{L}(\mathbf{p}, \mathbf{h}^*))] \qquad [6]$$

where the quantity between square brackets is usually called Negative Log-Likelihood (NLL). Given the expression above for $\mathcal{L}(\mathbf{p}, \mathbf{h})$, this would in practice amount to computing the set of parameters minimizing the sum of squared errors:

$$\mu_{post} = \mathrm{argmin}_{\mathbf{p}} \frac{1}{2\sigma_h^2} \sum_{i=1}^{nb_{wells}} (h_i^* - h(x_i, \mathbf{p}))^2 \qquad [7]$$

However, during the early development of this work, we noted that pursuing this approach would produce unrealistic results when applied to the head data available (these preliminary results are not shown in detail in this manuscript): indeed, running MODFLOW for the computed $\mu_{post}$ (which we recall is intended as the most likely value of the parameters) would predict that approximately 8-9% of grid cells would have calculated heads above the land surface, which is inconsistent with respect to field observations in the study area. To correct this problem, we modify our likelihood function in such a way that the drainage area anticipated by experts' opinion (based on previous "qualitative" knowledge about the peculiarities of the flow in study area) is also considered. More specifically, we add a factor to the likelihood function that models the fact that we expect that the most likely parameters should produce a drainage area $H_{PAS}$ of about $H_{PAS}^*$=1% and that in any case drainage areas outside the interval [0%, 2%] are unacceptable (which is reasonable given the observed occurrence of groundwater springs, fontanili, and inundated areas adjacent to the Ticino River). In other words, we treat the nominal drainage area $H_{PAS}^*$ (derived from experts' knowledge) as one additional experimental datum, that we model analogously to the well data: we assume it to be equal to the drainage area predicted by MODFLOW for the same unknown values of the parameters $\mathbf{p}_{true}$ plus a Gaussian random variable with zero mean and variance $\sigma_{H_{PAS}}^2$. In formulas:

$$H_{PAS}^* = H_{PAS}(\mathbf{p}_{true}) + \delta, \quad \delta \sim \mathcal{N}(0, \sigma_{H_{PAS}}^2) \tag{8}$$

where $\sigma_{H_{PAS}}^2$ is chosen as $\sigma_{H_{PAS}}^2$= 0.33%, such that $H_{PAS}(\mathbf{p}_{true}) = H_{PAS}^* - \delta$ exceeds [0%, 2%] with numerically zero probability. The resulting likelihood function is then

$$\mathcal{L}(\mathbf{p}, \mathbf{h}^*, H_{PAS}^*) = \left(\prod_{i=1}^{nb_{wells}} \frac{1}{\sigma_h \sqrt{2\pi}} e^{-\frac{(h(x_i, \mathbf{p}) - h_i^*)^2}{2\sigma_h^2}}\right) \frac{1}{\sqrt{2\pi}\sigma_{H_{PAS}}} e^{-\frac{(H_{PAS}(\mathbf{p}) - H_{PAS}^*)^2}{2\sigma_{H_{PAS}}^2}} \tag{9}$$

and repeating the same procedure described above for computing $\mu_{post}$ boils down to computing

$$\mu_{post} = \text{argmin}_\mathbf{p}[-\log(\mathcal{L}(\mathbf{p}, \mathbf{h}^*, H_{PAS}^*))] = \text{argmin}_\mathbf{p} NLL_{joint}(\mathbf{p}, \mathbf{h}^*, H_{PAS}^*) \tag{10}$$

where the new NLL function is defined as:

$$NLL_{joint}(\mathbf{p}, \mathbf{h}^*, H_{PAS}^*) = \frac{1}{2\sigma_h^2} \sum_{i=1}^{nb_{wells}} (h(x_i, \mathbf{p}) - h_i^*)^2 + \frac{1}{2\sigma_{H_{PAS}}^2}(H_{PAS}(\mathbf{p}) - H_{PAS}^*)^2 \\ + nb_{wells}\log(\sigma_h) + \log(\sigma_{H_{PAS}}) + \frac{nb_{wells}+1}{2}\log 2\pi \tag{11}$$

Note that $\sigma_h$ is also unknown and needs to be determined. While one way would be to simultaneously minimize $NLL_{joint}$ for the parameters $\mathbf{p}$ and $\sigma_h$, this method is not very robust numerically; therefore, we consider a two-step procedure in which we minimize $NLL_{joint}$ over a range of $\sigma_h$ values and finally select the combination of $\mathbf{p}$ and $\sigma_h$ which delivers the overall smallest $NLL_{joint}$. In doing so, for each fixed value of $\sigma_h$, the minimization with respect to the parameters is performed with a "composite method": first we evaluate $NLL_{joint}$ over a predefined cartesian grid of parameter values encompassing the uncertain ranges of the parameter values from Table 1; subsequently, the three parameter sets from the parameter grid with the lowest $NLL_{joint}$ values are used as starting points for a derivative-free optimization algorithm (simplex method, also called Nelder-Mead) to determine a refined optimal set of input parameters, and finally the overall best result out of the three is selected. More in details, the Cartesian grid is obtained by taking combination of 20 values of $R_{irrig}$, 15 values of $K_{zone1}$ and $K_{zone3}$, and 4 values of $K_{zone2}$ sampled logarithmically across the parameter range for a total of 6300 input parameter combinations across the multidimensional parameter grid (note that the logarithmic sampling is only used to consider also values towards the extrema of the intervals of $K_{zone1}$, $K_{zone2}$, $K_{zone3}$ as starting points of the optimization; this apparent mismatch with the pdf of the parameters does not impact the UQ analysis, since we do not compute statistical quantities out of the results of the optimization per se). The different numbers of sampling values for each parameter reflect their impact on the outputs of the model, as assessed in Baker et al. (2022). Moreover, as already mentioned, values of the Cartesian grid that do not respect the physical ordering $K_{zone3} \geq K_{zone2} \geq K_{zone1}$ were discarded.

### 3.3 Estimating the covariance matrix of the Gaussian approximation of the posterior pdf

After having computed the center of the approximate posterior pdf, i.e., the nominal value of the parameters, we now determine the covariance matrix of the posterior pdf, $\Sigma_{post}$, which quantifies the remaining uncertainty in the parameter values. More precisely, its diagonal entries are the variances of the parameters after the inversion, and the off-diagonal entries are the covariances between parameters. It can be shown that $\Sigma_{post}$ can be computed as follows:

$$\Sigma_{post} = \text{Hess}^{-1}[-\log \rho_{post}(\mu_{post})] \qquad [12]$$

i.e., the inverse of the Hessian of the negative log posterior distribution evaluated at its center, $\mu_{post}$, which is now known. Just like in the previous section, since $C$ and $\rho_{prior}$ are constant in our case, we can replace $-\log(\rho_{post})$ in the equation above with $NLL_{joint}$. Moreover, instead of computing directly the Hessian of the $NLL_{joint}$ we employ certain classical approximated formulas detailed in the following (Bui-Thanh et al., 2013; Piazzola et al., 2021; Nocedal et al., 1999), which require computing two easier objects only, namely the Jacobian of the model responses at $\mu_{post}$, i.e., the matrices of partial derivatives of the model responses at $\mu_{post}$ with respect to the uncertain parameters, both head measurements and surface flooding, called $J_h$ and $J_{HPAS}$ respectively. More in details, $J_h$ is a matrix with $nb_{wells}$ rows and 4 columns, defined as:

$$[J_h]_{l,j} = \frac{\partial h(x_l, \mu_{post})}{\partial p_j} \qquad [13]$$

and similarly for $J_{HPAS}$:

$$[J_{H_{PAS}}]_j = \frac{\partial H_{PAS}(\mu_{post})}{\partial p_j} \qquad [14]$$

To compute the entries of these matrices, we employ a forward finite difference scheme, centered at $\mu_{post}$ and with step $\Delta_j$ along each parameter which is proportional to the nominal value of that parameter (i.e., $\Delta_j = 0.02\% \mu_{post,j}$, the value 0.02% having been selected by a convergence study of the values of the partial derivatives at some pilot wells). For instance, denoting the components of $\mu_{post}$ as $\mu_{post} = [K_{zone1,post}, K_{zone2,post}, K_{zone3,post}, R_{irrig,post}]$, we have

$$[J_h]_{j,2} = \frac{\partial h(x_j, \mu_{post})}{\partial K_{zone2}} \approx \frac{h(x_j, [K_{zone1,post}, K_{zone2,post} + \Delta_{zone2}, K_{zone3,post}, R_{irrig,post}]) - h(x_j, \mu_{post})}{\Delta_2}, \Delta_2 = 2 \times 10^{-4} \times K_{zone2,post} \qquad [15]$$

Using these Jacobian matrices, the Hessian matrix can be approximated by the above-mentioned formula, which reads:

$$\text{Hess}[-\log \rho_{post}(\mu_{post})] \approx \frac{1}{\sigma_h^2} J_h^T J_h + \frac{1}{\sigma_{H_{PAS}}^2} J_{H_{PAS}}^T J_{H_{PAS}} \qquad [16]$$

and finally, $\Sigma_{post}$ is obtained by inverting the Hessian matrix just computed, see Eq. 12. Note that since the Hessian is computed at the minimum of $NLL_{joint}$, it is expected to be positive definite (i.e., its eigenvalues are expected to be positive), and therefore its inverse is expected to be positive definite as well, which is a property that must be satisfied by a covariance matrix. For a well-peaked posterior, $NLL_{joint}$ will have a narrow minimum, therefore the eigenvalues of Hess will be positive and large. Upon inverting Hess, we will then get that $\Sigma_{post}$ is a matrix with small diagonal entries, which means that the residual uncertainty on the values of the parameters is small. Conversely, a posterior that has a smeared peak will eventually lead to large diagonal entries in $\Sigma_{post}$, i.e., to a large residual uncertainty.

### 3.4 Identifiability analysis

As just motivated, the procedure outlined above, and in particular the assumption that the posterior pdf can be approximated by a Gaussian, works well if the posterior pdf is symmetric, unimodal and "well-peaked", which is typically true if the available data are "enough" and "not too noisy". To assess the validity of these assumptions, we perform an "identifiability analysis". Synthetic head data were generated by running the groundwater flow model using a known set of parameter values (see Table 1, row 2); Gaussian noise was then added to the synthetic head data, with a mean of 0 m and standard deviations ($\sigma_h$) of 0.25 m, 0.5 m, 1.0 m, 2.0 m, 3.0 m and 4.0 m. For each $\sigma_h$ value, 5 different sets of Gaussian noise were generated (for a total of 30 synthetic data sets). The procedure (minimization of $NLL_{joint}$ and computation of the covariance matrix) was then repeated for each set of Gaussian noises and the quality of the results (in terms of position and standard deviation of the posterior pdfs) was then compared to the known set of parameter values. Ideally, the following behaviors are expected from the posterior pdfs:
- they should be centered near the exact value of the corresponding parameters, such that the exact value is included in their support;
- the standard deviation of the pdf should not be too large (otherwise the posterior pdf would not be more informative than the prior about the value of the parameter) and in any case be small enough such that the parameters cannot be negative values (which would be unphysical) with practically zero probability.

### 3.5 Step 2: Forward UQ analysis of travel times

Upon verifying the validity of the gaussian approximation of the posterior, MODPATH version 7 (Pollock, 1988; Pollock, 1994; Pollock, 2016) was then used to perform particle tracking using the flow data generated by MODFLOW. Particles were placed in every fifth cell within the top layer of grid cells (for a total of 40090 particles) and tracked until they were discharged from the model. The porosity value of the aquifer was set to 0.2 based on data collected in the basin (Regione Lombardia, 2022). The tracking was repeated for 500 values of the uncertain parameters, generated according to their posterior distribution, to determine the effect of the residual parameter uncertainty on the estimated particle travel times; we point out that for each value of the uncertain parameters one needs first to solve the flow equation and then perform particle tracking. The entire procedure was repeated for 6 different posterior distributions, coming from one set of synthetic data for each level of Gaussian noise ($\sigma_h$ = 0.25 m, 0.5 m, 1.0 m, 2.0 m, 3.0 m and 4.0 m, for a total of 3000 particle tracking tests), to gain insight on the impact of the noise level on the results, and then finally applied to the posterior pdf obtained from the actual groundwater data (i.e. 500 further particle tracking tests). Summary statistics of the particle travel times were calculated for each of the 6 (synthetic) + 1 (observational) groups of particle tracking tests, to determine how the residual parameter uncertainty affects the estimated groundwater travel times. The examined summary statistics include calculating 25th, 50th, 75th, 90th, and 99th percentiles over the 40090 particles for each of the 500 simulations, and then taking the median, maximum and minimum of said percentiles over the 500 simulations at each level of noise and plotting these median, maximum and minimum values. Travel time distribution histograms and cumulative distributions were also plotted, as well as histograms of the 50th percentile travel times for each simulation, and the percentage of particles with travel times less than 25 years for each of the simulations. Prior to examining these summary statistics, particles with a travel time of exactly 0 were excluded since these represent particles that were placed in discharging cells (e.g., sinks such as river cells and spring cells) and so never entered the modeled groundwater system.

## 4. Results & Discussion

We are now ready to discuss the results of the different steps of our analysis: we first discuss in Section 4.1 the results obtained by applying the Bayesian inversion procedure to the synthetic head data sets, highlighting advantages and limitations of the gaussian approximation of the posterior pdf of the uncertain parameters of MODFLOW. This allows us to apply the same procedure to the real data sets with more confidence on the interpretation of the results (Section 4.2). Then, in Sections 4.3 and 4.4 we perform the forward UQ analysis for the travel times, propagating through MODPATH the uncertainty on the parameters as encoded in their posterior pdfs. The insights gained in Section 4.3 on synthetic data provide us with a deeper understanding of the results obtained on the real data in Section 4.4.

### 4.1 Identifiability analysis
The posterior pdfs obtained applying the Bayesian procedure with gaussian approximation for the different sets of synthetic head data considered are reported in Figure 2 (posterior pdfs of $K_{zone1}$ and $K_{zone2}$) and Figure 3 (posterior pdfs $K_{zone3}$ and $R_{irrig}$). Each panel shows the five posterior pdfs computed for the datasets at the same level of noise $\sigma_h$, and moving downward across panels in the same column shows the behavior of the posterior pdfs as the noise on the measurements increase. Furthermore, the black vertical solid line shows the exact value of the parameter (cf. Table 1, row 2), whereas the two dashed lines mark respectively half and twice the exact value. The green background highlights the support of the prior pdf. Several observations can be drawn from these figures:
- For small values of $\sigma_h$, the exact values of the parameters are quite accurately recovered, since the centers of the pdfs are all very close to the exact value. As $\sigma_h$ increases, the estimated parameter values drift to values that are further from the true (base case) parameter values. However, even at the largest levels of noise the estimated parameter values are still close to the exact values, most often within a factor of 2 (or 0.5) from the exact value, which we deem small enough for our purposes. A closer look at the values of the ratios between the estimated and exact parameter values is shown in Figure 4.
- In Figure 4 we also report the ratio between the estimated and exact values of the Gaussian noise applied to create the synthetic data (see discussion after Equation 11). This ratio is very close to 1 for every value of noise, which indicates that we can accurately estimate the amount of uncertainty in well observation data.
- $K_{zone3}$ is the parameter most inaccurately predicted at lower levels of noise in the synthetic data ($\leq 1.0$ m). This is likely because only a few observation wells are located within this region, affecting the reliability with which this parameter can be estimated. Meanwhile, at higher levels of noise (≥2.0 m) the recharge rate due to irrigation is typically the least accurately predicted. This might be due to a trade-off effect between $R_{irrig}$ and the values of K, since if $R_{irrig}$ increases, the conductivities might end up increasing accordingly to match the original well data - however, analysis of the correlations between the different K values and $R_{irrig}$ (not shown for brevity) did not support this intuition, and further investigations on this matter are left for a future work. $K_{zone1}$ is typically predicted most accurately regardless of the level of noise in the synthetic data, likely because the highest number of observation wells are in this area.
- On average, the standard deviations of the posterior pdfs also increase with the level of noise; moreover, as the level of noise increases, the standard deviations are less consistently estimated (i.e., the estimated standard deviations are not always similar for fixed large values of sigma). In particular, for larger values of noise it occasionally happens that the standard deviation of the posterior is so large that negative (hence unphysical) values of the parameters have non-zero probability - which suggests that results have to be taken cautiously for larger values of noise. A more quantitative evaluation of this issue is provided in Figure 5a, where we show the *coefficient of variation* of the posterior pdfs (ratio of their estimated standard deviation to their mean, ie. to the predicted parameter value): this coefficient increases with the noise level, as does its spread across data sets. The horizontal dashed line marks the threshold 0.5 (i.e., STD ≥ 0.5 mean, or equivalently mean ≤ 2 STD), above which the gaussian posterior can provide negative values of the parameters with a probability that is non-negligible from an engineering point of view (>2.5%): this is due to the well-known fact that for a Gaussian variable approximately 95% of the values are within the interval [mean - 2STD; mean + 2STD], which means that the mean must be larger than 2STD to have less than 2.5% probability that the parameter can assume negative values.
- The desired property that the exact value of the parameter is always included in the range of the pdf is unfortunately not valid, even for small values of noise. In the case of small values, this problem is mitigated by the fact that the predicted values of the parameters are in any case very accurately predicted, and the associated standard deviations are small (therefore, we are never committing a large mistake). In the case of larger noises, results are still somewhat encouraging since most often 2 or 3 pdfs out of 5 still encompass the exact value, which means that even poor data might still give somewhat reliable information.
- The range of the posterior pdf is often significantly smaller than the range of the prior pdf, which means that in general the procedure can be very effective in reducing the uncertainty on the values of the parameters. Figure 5b shows the ratio of the posterior variance to the prior variance of the parameters, to provide a quantitative insight on the variance reduction obtained after the inversion procedure. Only a few realizations show a ratio equal to or above 1, while most (106 out of 120) show the variance in the parameters has been reduced by at least a factor of 10.

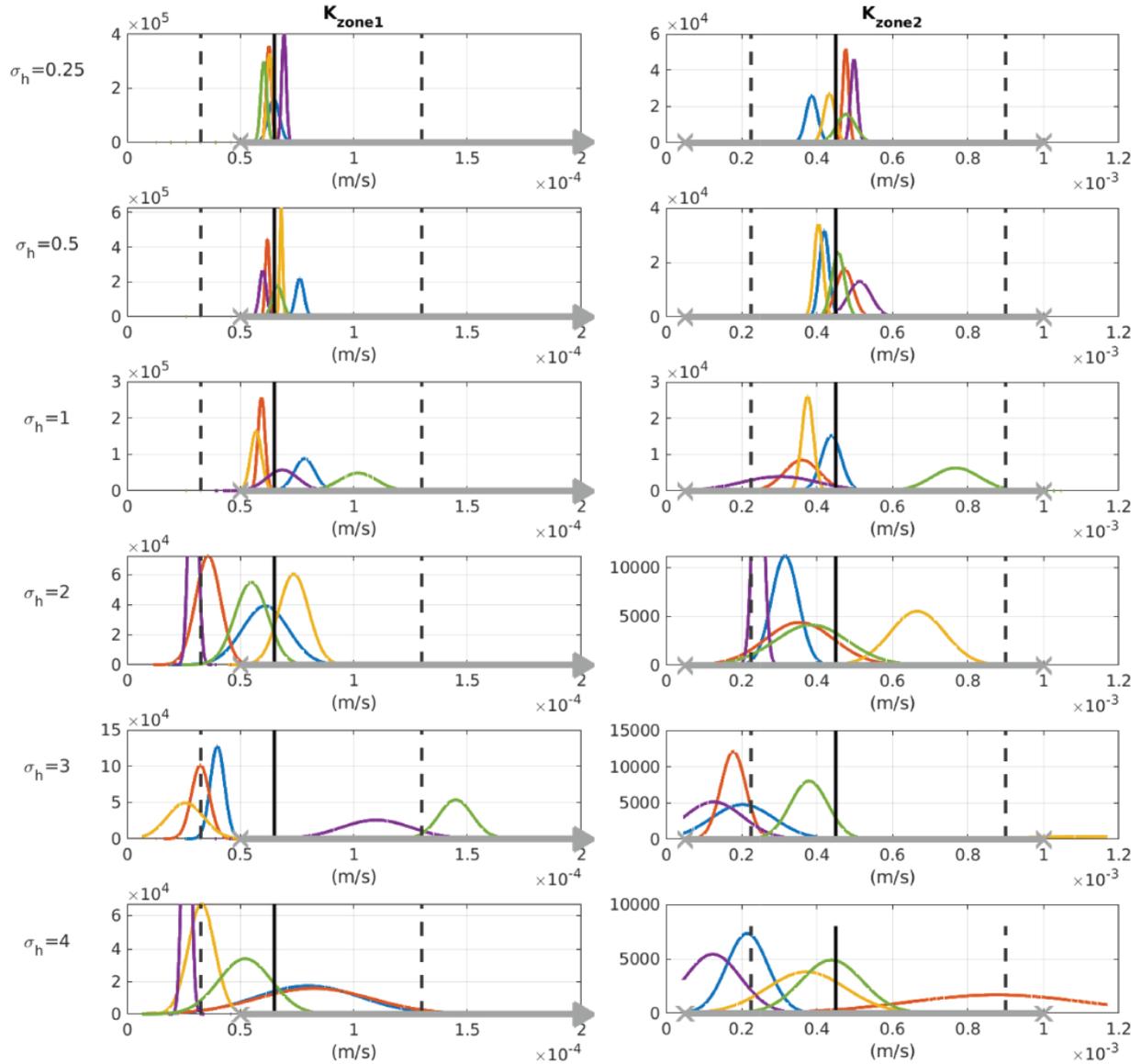

**Figure 2:** For a fixed row, each panel shows the five posterior pdfs computed for the datasets at the same level of noise, $\sigma_{h\_noise}$, for $K_{zone1}$ (left panel) and $K_{zone2}$ (right panel). Different rows report the results obtained as the noise on the measurements increases from $\sigma_h$=0.25 (top row) to $\sigma_h$=4 (bottom row). The black vertical solid line shows the exact value of the parameter (cf. Table 1, row 2), whereas the two dashed lines mark respectively half and twice the exact value. The thick gray line with "X" markers on the horizontal axis denotes the support of the prior pdf: for $K_{zone1}$, the right marker is replaced by an arrowhead, denoting that the prior support would extend further to the right but is not shown to maximize visibility. The unit of measure of the vertical axes is $(m/s)^{-1}$ (not shown in the panels for sake of readability). Note that the scale of the vertical axis is not constant across noise levels, to maximize visibility.

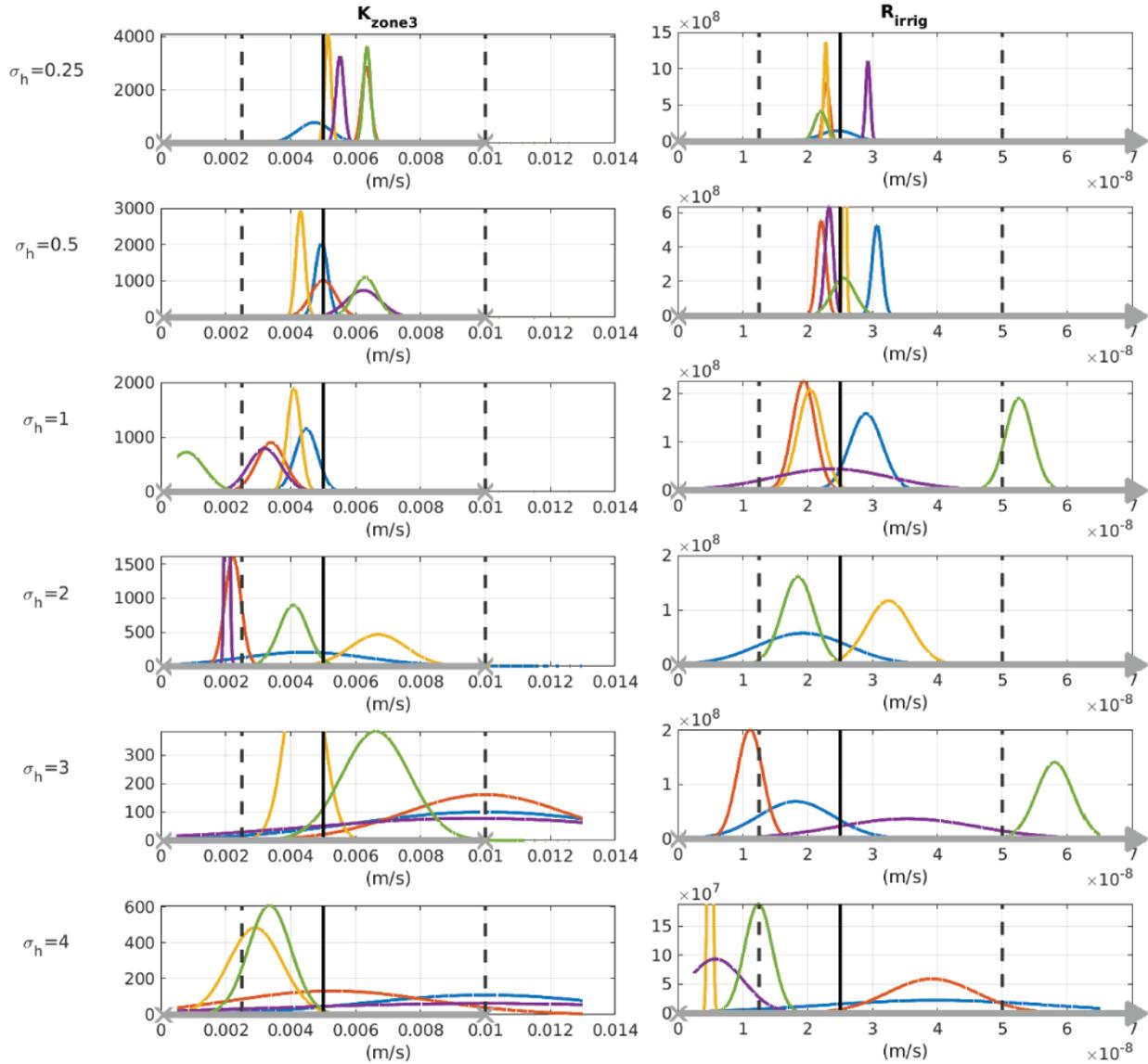

**Figure 3:** For a fixed row, each panel shows the 5 posterior pdfs computed for the datasets at the same level of noise $\sigma_{h\_noise}$, for $K_{zone3}$ (left panel) and $R_{irrig}$ (right panel). Different rows report the results obtained as the noise on the measurements increases from $\sigma_h=0.25$ (top row) to $\sigma_h=4$ (bottom row). The black vertical solid line shows the exact value of the parameter (cf. Table 1, row 2), whereas the two dashed lines mark respectively half and twice the exact value. The thick gray line with "X" markers on the horizontal axis denotes the support of the prior pdf: for $R_{irrig}$, the right marker is replaced by an arrowhead, denoting that the prior support would extend further to the right but is not shown to maximize visibility. The unit of measure of the vertical axes is (m/s)$^{-1}$ (not shown in the panels for sake of readability). Note that the scale of the vertical axis is not constant across noise levels, to maximize visibility.

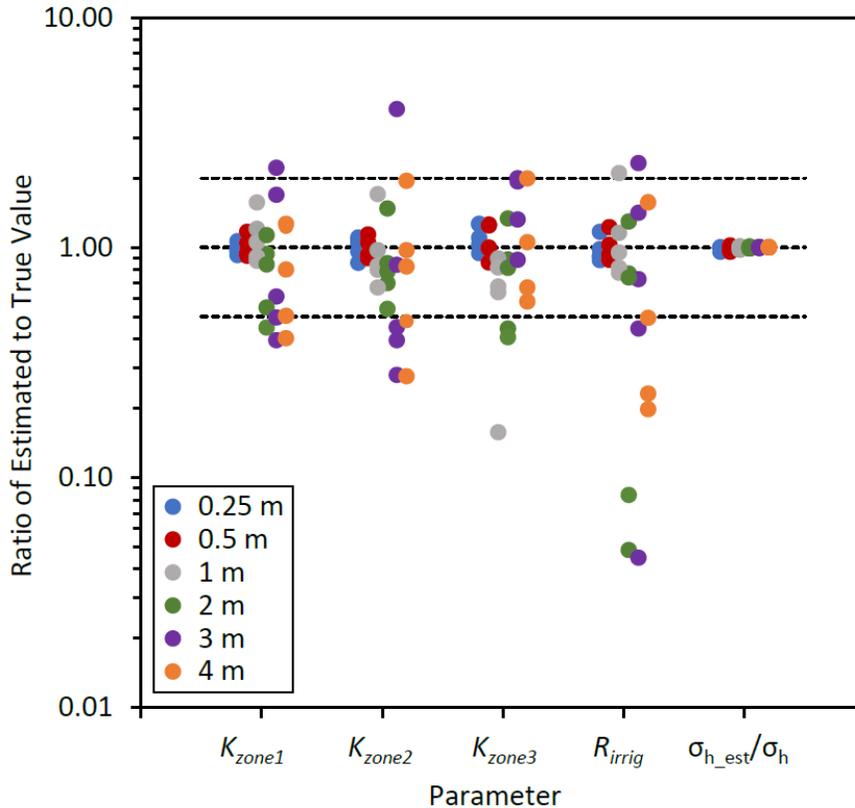

**Figure 4:** Jitter plot (with y-axis in log scale) of the ratios between the predicted and true parameter values for each parameter at different levels of Gaussian noise $\sigma_h$ (color of symbol). The three dashed lines show the location of ratios equal to 0.5 / 1 / 2 (from lower to upper lines). The last column shows the same information for the ratio between the estimated and true value of the artificial noise used to generate the synthetic heads.

All considered, the gaussian approximation of the posterior gives partially satisfactory results. It is quite effective in delivering approximated values of the parameters, which gives us confidence that the results obtained when running MODFLOW with them can be representative. However, it is less satisfactory in quantifying the residual uncertainty on the parameters especially if the noise on the data is too large, which means that forward UQ analysis might be biased (i.e., structurally underestimating or overestimating reality). However, even for large levels of noise the results can be good: for example, at 3 m and 4 m of noise the nominal value of the parameters is included in the predicted range of uncertainty 11 times out of 20 and 13 times out of 20, respectively. The lesson learnt is that results must be approached judiciously but are not to be entirely distrusted. It is particularly crucial to be able to estimate the level of noise affecting the data: in this respect, our results suggest that the employed algorithm is quite reliable. Algorithms that deliver an approximation of the posterior pdf without resorting to an assumption of gaussian approximability are available in the literature (Markov-Chain Monte Carlo algorithms, see, e.g., Brooks et al, 2011) and might help in providing posterior pdfs whose range "always" includes the exact parameter values. However, these algorithms are typically computationally expensive, require adjusting a number of tuning parameters and are also not immune to data quality issues. The assessment of their performance in the context of our problem is thus left to future works.

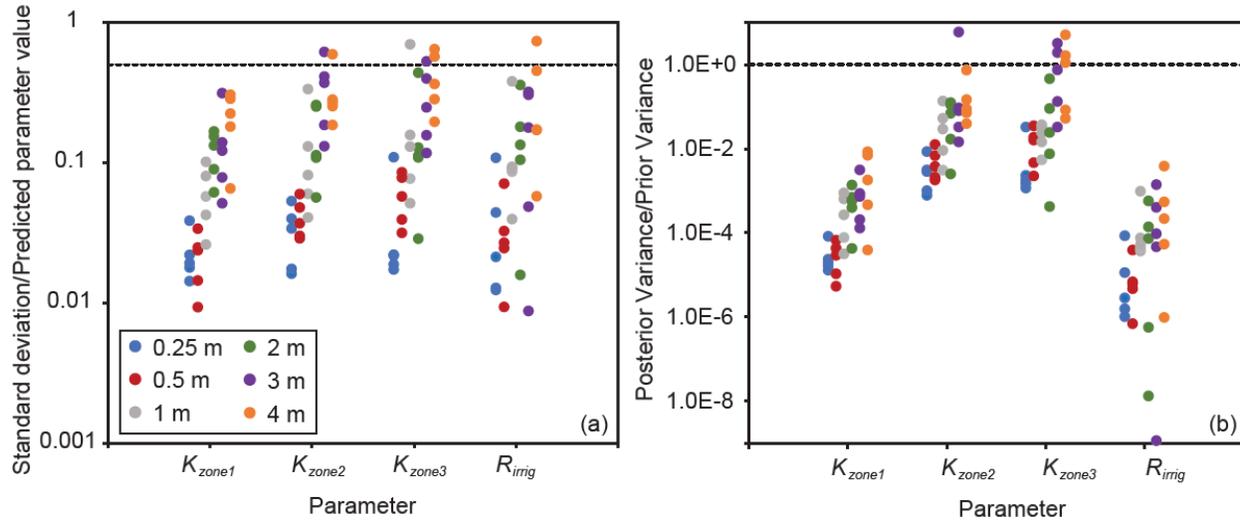

**Figure 5.** (a) Ratio of the estimated standard deviation of the parameter value to the predicted parameter value for each model parameter at each level of Gaussian noise (color of symbol). The black dashed line is where the standard deviation is half the value of the predicted parameter value. (b) Ratio of the posterior variance to the prior variance of the parameters at each level of Gaussian noise. The black dashed line indicates where the posterior and prior variances are equal.

### 4.2 Inversion of Real Data

The insights gained from analysis of the optimization procedure for the synthetic data can then be applied to the real observation well data from the study area. When the NLL functional is minimized using the observed groundwater head data and a $\sigma_{H_{PAS}}$ of 0.33%, $\sigma_h$ is equal to 3.60 m and the optimized parameter values and associated posterior standard deviations are shown in lines 3 and 4 of Table 1. Such a value of $\sigma_h$ (comparable to the RMSE of the heads of 3.64 m) is a realistic amount of uncertainty in our observed groundwater heads given that they were measured over a span of almost a month, some levels may not have fully recovered from pumping prior to measurement, and because the groundwater flow model is highly simplified compared to the natural groundwater system.

The results shown in Figure 2 from the synthetic data show that even when data with $\sigma_h$ values of 3 to 4 m are used, the parameter values can be estimated within half an order of magnitude or less of their true value, which is a large improvement from the initial parameter uncertainty which can often span multiple orders of magnitude for input parameters such as aquifer hydraulic conductivity and recharge. The resulting percentage of model grid cells with calculated heads above the land surface is 1.26%, which is reasonable given the abundance of springs in the study area that may occur in slightly different locations in the model domain than in reality and given the wetland areas adjacent to certain sections of the Ticino River. The modeled flux rate of groundwater from the springs (79.1 l/s) and the groundwater flux rate into the Ticino River (1.38×10$^{-4}$ m$^3$/s/m) are also reasonable (Baker et al., 2022) and in accordance with the results of previous studies (De Luca et al., 2014; Balestrini et al. 2021; Musacchio et al., 2021), supporting the accuracy of the groundwater flow model.

The posterior pdfs of the uncertain parameters after inversion are shown in Figure 6. The pdfs are such that the probability of having negative values of the parameters is essentially zero. The standard deviations of the pdfs are to be contrasted with the last two rows of Figures 2 and 3, as similarly the coefficients of variation of these pdfs (see Table 1, row 5) should be compared with the orange and purple dots reported in Figure 5(a): upon inspection, the posterior pdfs based on the real data are comparable with the narrower pdfs for synthetic data at similar values of noise.

Notice though that for such levels of $\sigma_h$, Figures 2 and 3 also point out that we might be in the situation such that the exact parameters values are not included in the range of the posterior, such that we cannot blindly rely on the ranges of the parameters suggested by the posterior pdfs.

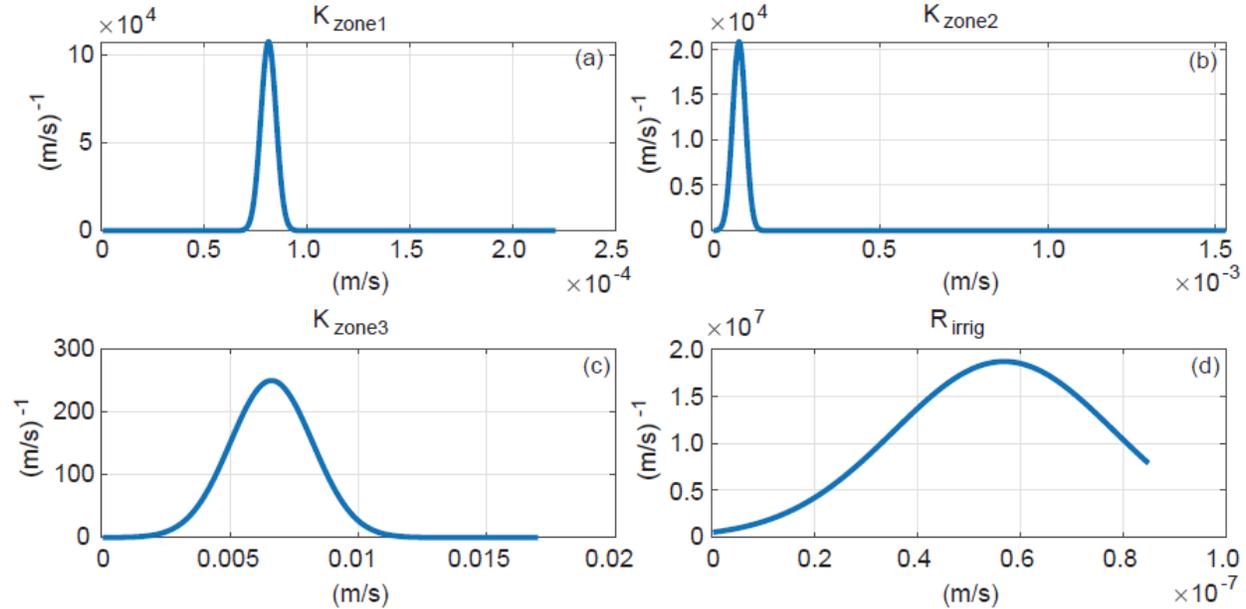

**Figure 6.** Posterior pdfs of the uncertain parameters after inversion based on the observed well data.

### 4.3 Forward Uncertainty Quantification of Particle Travel Times for synthetic data

The estimated posterior pdfs obtained in Section 4.1 can then be used to understand how the residual uncertainty on the parameter values impacts the robustness of the calculated distributions of groundwater travel times. Note that even if such pdfs are not always optimal in terms of "consistency with the true parameters", it is nonetheless instructive to understand how the uncertainty in the random parameters propagates to the travel times, e.g., whether uncertainty gets amplified, if the symmetry in the pdf of the parameters is lost, etc.

Travel time distributions were calculated in MODPATH, using first the base values of the parameters used to generate the synthetic data (see Table 1) and then considering 500 parameter sets for each level of Gaussian noise (3000 travel time distributions in total). Each of the 500 parameter sets were randomly generated from the normal, multivariate Gaussian distribution using the optimized parameter values and the corresponding posterior covariance matrices.

For the simulation with the base values, the 25th, 50th, 75th, 90th, 99th percentiles of the travel time distributions are reported in Table 2. For the other simulations, Figure 7 reports for each of the six levels of noise the histograms of travel time distributions for all of the 500 sets of parameter values that were considered: the results show that the particle travel times consistently follow a seemingly exponential distribution across the 500 realizations at all levels of Gaussian noise, with many particles having short travel times of only a few years and most having travel times less than about 20-25 years. As expected, as the amount of Gaussian noise applied to the synthetic head data increases from 0.25 m to 4.0 m, the travel time distributions show a larger variability across the 500 realizations, particularly in the range of Gaussian noise from 1.0 to 4.0 m, as seen in the cumulative distributions plotted in Figure 8, indicating that the more noise in the head data the more uncertainty on the travel time distributions. The horizontal and vertical dashed lines in Figure 8 indicate the 50th percentile of the travel times and the percentage of particles with travel times less than 25 years (approximately the 75th-80th percentile), respectively. The histograms of these cross-sections are plotted in Figures 9 and 10.

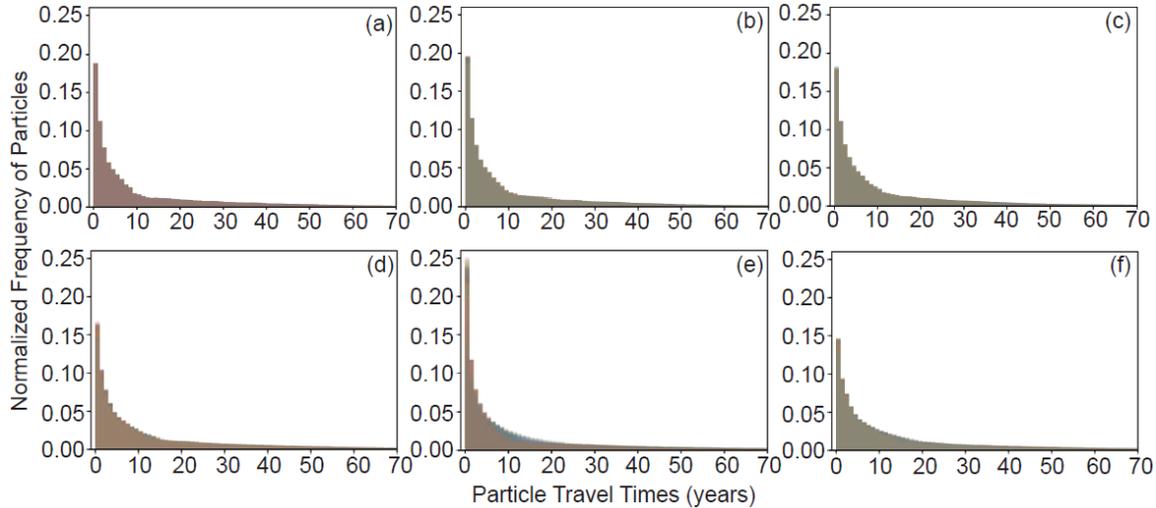

**Figure 7**. Travel time distributions for the 40090 particles released in the model configurations optimized using the synthetic groundwater head data with random Gaussian noise of (a) 0.25 m, (b) 0.5 m, (c) 1.0 m, (d) 2.0 m, (e) 3.0 m, and (f) 4.0 m. We report in each panel 500 travel time distributions, each corresponding to one of the 500 sets of input parameters generated from their gaussian posterior density.

**Table 2**. Particle travel times in years for the base case model run that uses the parameter values in line 2 of Table 1.

| Travel time percentiles over the 40090 particles | | | | |
|---|---|---|---|---|
| 25% | 50% | 75% | 90% | 99% |
| 1.73 | 5.55 | 18.16 | 42.68 | 136.09 |

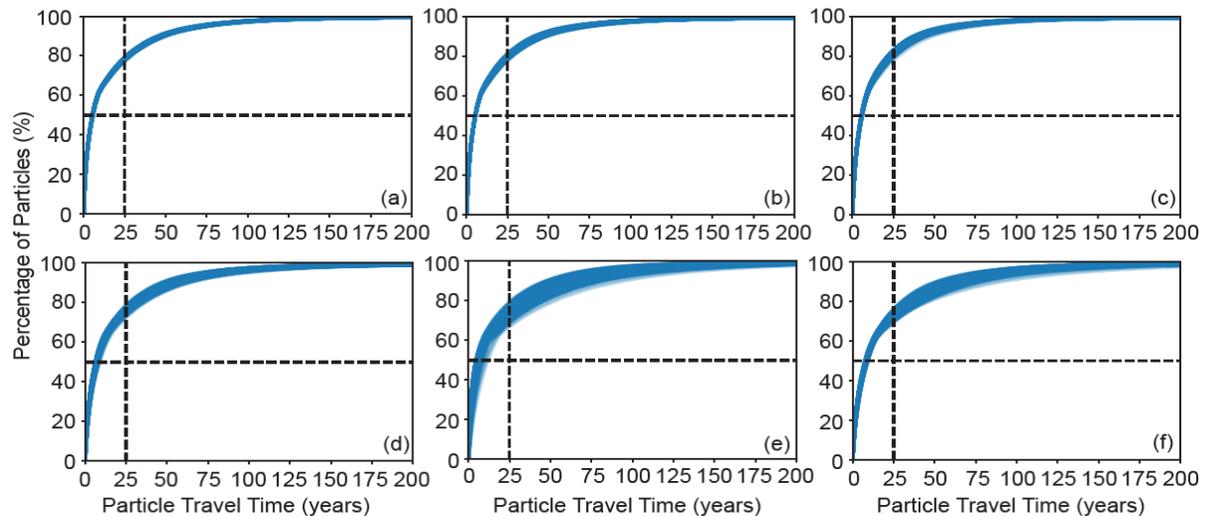

**Figure 8**. Cumulative travel time distributions for the 40090 particles released in the model configurations optimized using the synthetic groundwater head data with random Gaussian noise of (a) 0.25 m, (b) 0.5 m, (c) 1.0 m, (d) 2.0 m, (e) 3.0 m, and (f) 4.0 m. The sets of travel time distributions were each generated using 500 sets of input parameters generated from their mean value and covariance. The horizontal dashed line crosses at the 50th

percentile travel times, with the distribution of these travel times shown in Figure 9. The vertical dashed line crosses at a travel time of 25 years, with the distribution of these travel times shown in Figure 10.

In detail, the histograms of the 50th percentile travel times are plotted in Figure 9 (horizontal cross-sections of Figure 8) and can be compared with the corresponding base case value for the same percentile reported in Table 2 (5.55 years). Ideally, we wish that all the histograms in Figure 9 are centered at 5.55 years (or at least that the histograms are supported over an interval that includes 5.55 years) but this depends on the extent to which the posterior pdfs of the parameters are centered close to the exact values of the parameters. Inspection of Figure 9 reveals that for 0.25, 0.5, and 1.0 m of noise in the synthetic data we are in the ideal situation: at these levels of noise the 50th percentiles of the travel times are indeed about 5.2-5.8 years, with similarly symmetric shaped histograms of sets of 500 simulations. Meanwhile, at higher levels of Gaussian noise (2.0, 3.0, 4.0 m) the consistency with the base case is lost. In particular, for 2 m and 4 m of Gaussian noise (Figure 9d and 9f, respectively) the histograms show that the 50th percentile is overestimated for all the 500 simulations, while for 3m of Gaussian noise (Figure 9e) the histogram of the 50th percentile has a maximum close to 5.55 years but is skewed towards larger values instead of being symmetric as at the smaller values of noise. These results are consistent with Table 3, where we report the ratio of the estimated to true parameter values for the six cases considered in Figure 9: while all ratios are between 0.5 and 2 (as in Figures 2-3), in the three cases of smaller Gaussian noise such ratios are much closer to 1 (perfect estimate) than at the larger three levels of noise. Note in particular that for the cases of 2 m and 4 m of noise, all the parameters are underestimated, which means that the three predicted permeabilities are too smaller, implying longer travel times for the particles; conversely in the case of 3 m of noise, $K_{zone3}$ (permeability of the zone closest to the river, see Figure 1d) is overestimated, which implies faster travel through zone 3, which compensates for the slower travel in zones 1 and 2, such that the overall travel times are coincidentally accurate. It is useful to highlight that a factor of at most two in underestimating the parameters translates into a 50th percentile of the travel times which is also off by a factor two, which means that errors are still acceptable for engineering evaluations.

**Table 3**. Ratios of the estimated to true parameter values. The true parameter values are in line 2 of Table 1. The estimated parameter values are obtained from the NLL optimizations using the synthetic head data at each level of Gaussian noise, $\sigma_h$.

|  | Parameters | | | |
| --- | --- | --- | --- | --- |
| $\sigma_h$ (m) | $K_{zone1}$ | $K_{zone2}$ | $K_{zone3}$ | $R_{irrig}$ |
| 0.25 | 0.93 | 1.06 | 1.27 | 0.88 |
| 0.5 | 1.02 | 1.01 | 1.26 | 1.02 |
| 1.0 | 1.20 | 0.97 | 0.90 | 1.16 |
| 2.0 | 0.84 | 0.86 | 0.82 | 0.74 |
| 3.0 | 0.61 | 0.45 | 2.00 | 0.73 |
| 4.0 | 0.80 | 0.97 | 0.67 | 0.49 |

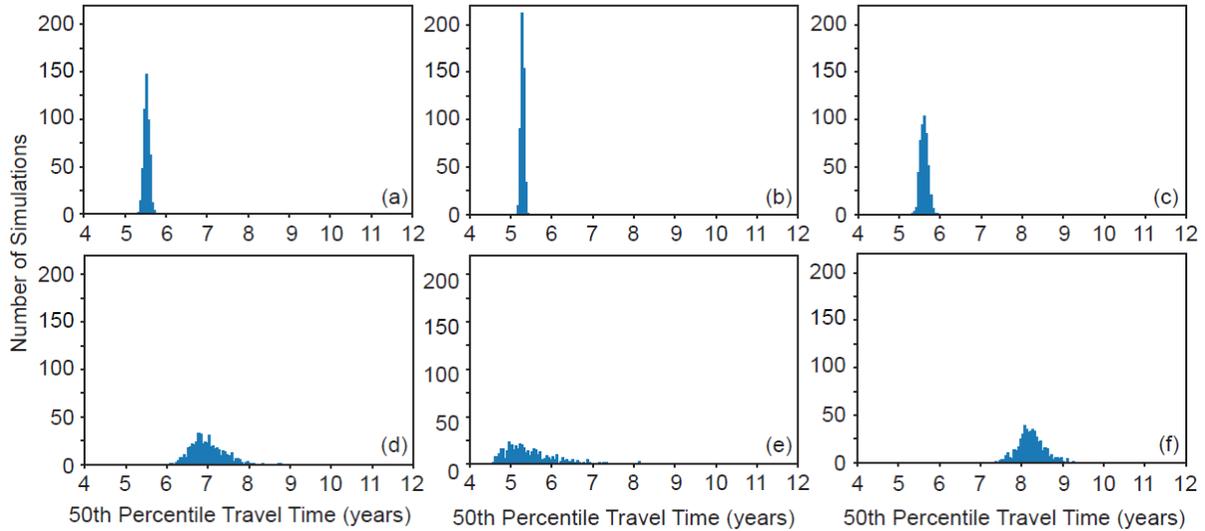

**Figure 9**. Histograms of the 50th percentile travel times over the 500 simulations for each synthetic data set (horizontal cross-sections of Figure 8) using fixed bin widths of 0.05%. Synthetic data sets were created by adding (a) 0.25 m, (b) 0.5 m, (c) 1.0 m, (d) 2.0 m, (e) 3.0 m, and (f) 4.0 m of Gaussian noise added to the base case heads.

Figure 10 contains histograms of the percent of particles with travel times less than 25 years for each of the 500 simulations at each level of Gaussian noise (vertical cross-sections of Figure 8). These plots indicate that about three-quarters of the particles have travel times less than 25 years. At lower levels of noise (0.25 to 1.0 m), about 77-83% of particles exit the model through rivers or springs within 25 years of recharging the aquifer. As the noise increases to 2.0 to 4.0 m, the range of this percentage decreases while spanning a broader group of values, such that anywhere from 68-79% of particles exit the model within 25 years depending on the simulation. Moreover, as the level of noise increases, the histograms of the percent of particles with travel times less than 25 years become less peaked, depending on the estimated standard deviations and covariances between the parameters, indicating more uncertainty in the predicted travel times when the amount of Gaussian noise in the synthetic data is greater.

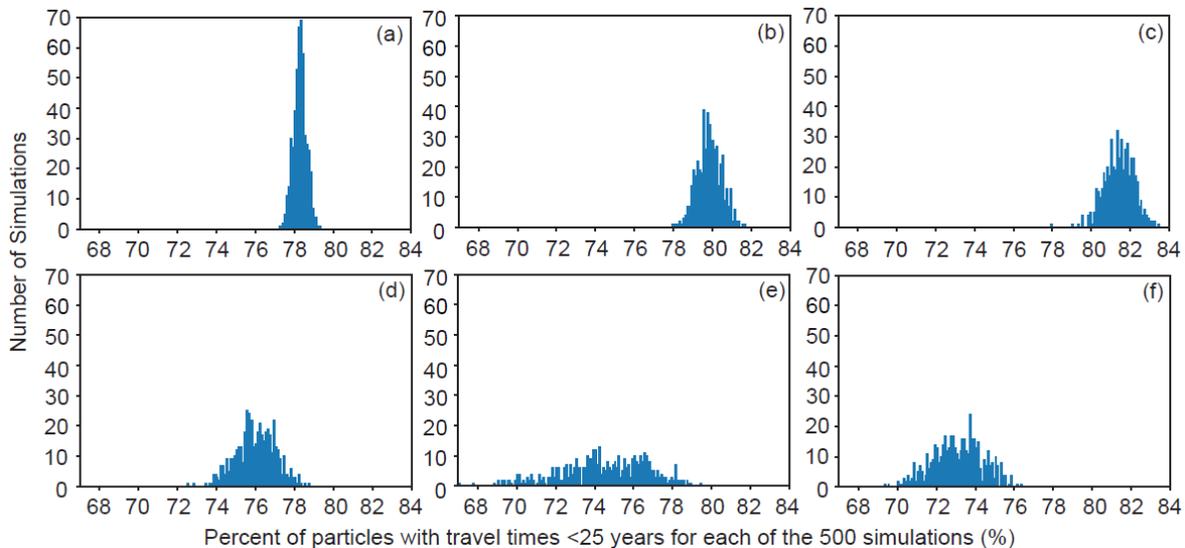

**Figure 10**. Histograms of the percentage of particles with travel times less than 25 years for each of the 500 simulations (vertical cross-sections of Figure 8) using fixed bin widths of 0.1%. (a) 0.25 m, (b) 0.5 m, (c) 1 m, (d) 2

m, (e) 3 m, and (f) 4 m of Gaussian noise was added to the synthetic head data used to determine the optimal parameter values and their estimated covariances.

Going further in the analysis, Figure 11 extends the results of Figure 9 and shows the median, maximum and minimum across the 500 computed travel time distributions of the 25th, 50th, 75th, 90th and 99th percentile of travel times (i.e., we show results for additional horizontal cuts of Figure 8). This figure shows that higher levels of Gaussian noise in the well data results in more variability in the predicted travel times, especially at the larger percentiles, whose prediction is thus less reliable. Indeed, the median, maximum and minimum of all the analyzed percentiles over the 500 travel time distributions are similar for up to 1.0 m of applied Gaussian noise; then the values begin to differ and to be more spread out. Had travel time distributions been computed with the prior pdf, there would be even more uncertainty in the travel times. Instead, by reducing the uncertainty in the parameters during the inversion, especially for $K_{zone1}$ and $R_{irrig}$, the uncertainty in the travel times have also been reduced. Reducing the uncertainty of $K_{zone1}$ was especially helpful because this zone has the lowest hydraulic conductivity values and so acts to slow down the particles traveling through it.

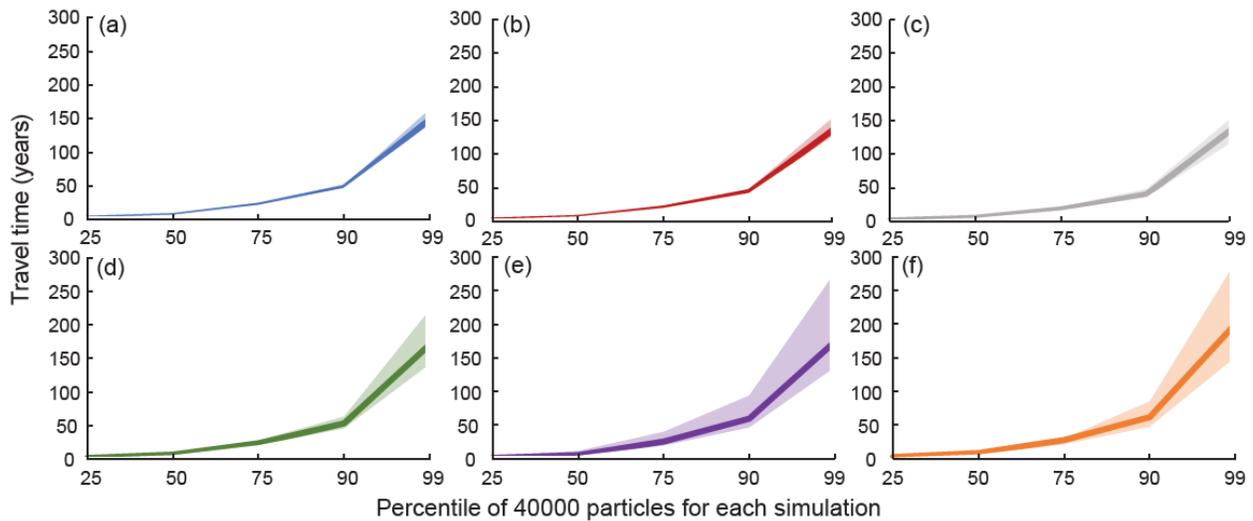

**Figure 11**. Percentiles of the particle travel times for the 500 simulations at each level of noise. The solid line in the envelope represents the median travel time at each of the percentiles for the 500 simulations while the boundaries of the envelope are the maximum and minimum travel times for each percentile over the 500 simulations.

### 4.4 Uncertainty Quantification of travel times for Real Data

When the nominal parameter values that were predicted from the observed head data and expected drainage area (i.e., those obtained upon minimizing the NLL functional, see Sections 2.4 and 3.4) are used to calculate the groundwater travel time distributions, the travel times of the 40090 particles for the recharge due to precipitation and rice field irrigation have an average of 10.9 years, with a 50% percentile of travel times of about 3.9 years. Additional travel time percentile data for this simulation are in the last row of Table 4. The travel time of every particle calculated using the nominal parameter values are shown in Figure 12a; particles that recharge the aquifer in locations closer to the center of the valley, where the stream is located, tend to have shorter travel times on the order of a few months to a few years, while particles that recharge the aquifer closer to the eastern and western boundaries of the basin have longer travel times (on the order of 10-50 years) since the particles must both travel a greater distance and through Zone 1 which has a lower hydraulic conductivity. However, there are some particles with shorter travel times near the edges of the basin, which are particles discharged through drain boundaries (i.e., springs).

Table 4. Median, maximum, and minimum particle travel times across various percentiles for 500 model runs and for the NLL optimized set of parameters where the optimized parameter values were estimated using the observed groundwater head data. The sets of parameter values for the 500 runs were generated according to the calculated covariances of the optimized parameter values.

|  | Travel time (years) | Percentiles over the 40090 particles ||||| 
|---|---|---|---|---|---|---|
|  |  | 25% | 50% | 75% | 90% | 99% |
| Summary statistics across 500 model simulations | Max. | 1.53 | 5.77 | 24.68 | 49.45 | 131.49 |
|  | Min. | 0.94 | 3.03 | 9.56 | 23.02 | 78.55 |
|  | Median | 1.21 | 4.09 | 13.66 | 30.67 | 95.99 |
| Statistics of the model run with optimized parameters based on observed head data |  | 1.23 | 3.85 | 12.17 | 28.24 | 90.37 |

To quantify the uncertainty on these nominal predictions, we perform once more the procedure employed in the previous section, i.e., the estimated covariance matrices of the nominal values of the parameters are used to generate a group of 500 model runs, whose travel time cumulative distributions are shown in Figure 12b. The mean over the 500 model runs of the average travel time of the 40090 particles is 11.9 years while the median across the 500 simulations of the 50th percentile of the particle travel times is 4.1 years. The maximum and minimum across the 500 simulations for this percentile (as well as for the 25th, 75th, 90th, 99th) were also calculated, see Table 4. The maximum and minimum across the simulations are similar up to the 50th percentile, while their spread is larger at the higher percentiles, especially past the 90th percentile, as shown also in Figure 12c. Figures 12d and 12e show the 50th percentile of the travel times (analogous to Figure 9) and the percentage of particles with travel times less than 25 years for the set of 500 simulations (analogous to Figure 10). The histogram of the 50th percentile travel times is fairly symmetric, with median travel times typically between 3-5 years, but with a slight tail in the direction of longer median travel times up to 5.8 years. Meanwhile the histogram of the percentage of particles with travel times less than 25 years is skewed to the left, with 85-91% of particles typically exiting the model within 25 years, but with some simulations having less than 80% of particles exiting in 25 years, as shown in Figure 12e.

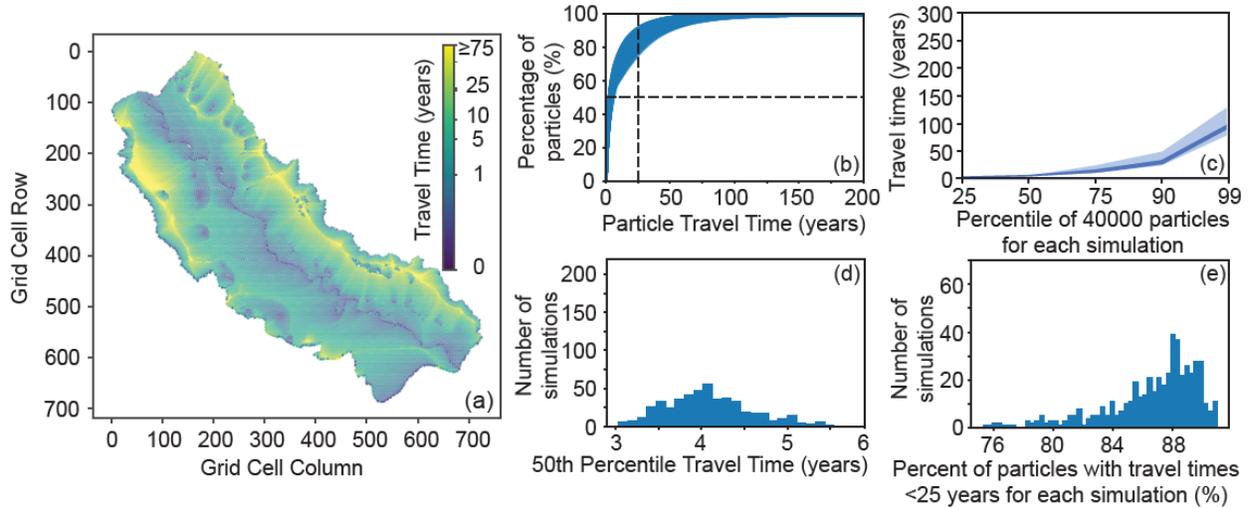

**Figure 12**. Travel time results when the observed head data are used to predict the model parameter values using the NLL optimization procedure. (a) Particle travel times according to starting locations using the optimized parameter values. (b) Travel time distributions for 500 different parameter sets generated from the estimated parameter standard deviations and covariance matrix. (c) Maximum, minimum, and median travel times of the 500 simulations across various percentiles. (d) 50th percentile travel times of the particles for each of the 500 simulations. (e) Percent of particles with travel times less than 25 years for each of the 500 simulations.

Since the real observation well data is estimated to have an uncertainty of 3.6 m, the results from the synthetic simulations using 3-4 m of Gaussian noise are most appropriate for comparison. Therefore, the estimates of the true parameter values may contain some error, though most likely within a factor of 2 (Figures 2-4), and we might be in the situation such that the exact parameters values are not included in the range of the posterior, such that travel time predictions must be interpreted cautiously: consistent underestimation of the parameter values might lead to overestimation of the travel times, overestimation might lead to underestimation of the travel times, and finally if some parameters are overestimated and others underestimated, this can sometimes result in seemingly accurate travel time predictions. While it is impossible to know whether the parameter values estimated using the real data are overestimated or underestimated, the results from the synthetic data demonstrate that even with errors in the estimated parameter values, the estimated travel times are still accurate within less than a factor of 2. Using the method detailed above we are able to quantify the amount of uncertainty in the estimated travel times induced by the residual uncertainty after inversion, and ultimately see that this uncertainty does not affect the conclusion that groundwater travel times in the lower Ticino basin are relatively short. The travel time distributions represent a first estimate for the study area since the conditions modeled are only for the period of August/September 2014. However, the same method can be applied to future transient groundwater flow models that encompass longer time periods with a greater range of climatic and irrigation conditions.

## 5. Conclusions

This work deals with the simulation of groundwater transport in the aquifer of the Ticino river basin. As is common in groundwater flow simulations, the input parameters are uncertain. In this study we perform Uncertainty Quantification (UQ) analysis by incorporating suitable error metrics into the Negative Log Likelihood (NLL) functional (Eq. 11). We demonstrate how the NLL functional can be used to incorporate multiple error metrics into a single optimization function. The resulting optimized parameter values can then be used to calculate the covariance matrix of the parameters. Although the RMSE and the percentage of modeled grid cells with heads above the land surface ($H_{PAS}$) were the error metrics incorporated into the NLL functional in this study, different or additional error metrics could be used in other studies depending on the most informative error metrics for the given field site. For example, the measured and modeled groundwater flux rate into a river could be substituted for $H_{PAS}^*$ and $H_{PAS}$, and $\sigma_{H_{PAS}}$ would instead be a value that defines the range of plausible groundwater flux values to the river. The values of the optimized parameter standard deviations and their related covariance matrices can then be used to assess the potential variability in the calculated groundwater travel time distributions due to this remaining uncertainty. By first applying such a method to synthetic data created by adding Gaussian noise to model generated head data, the modeler can assess the given model's ability to make accurate predictions under their specific set of conditions and uncertainties. In this work the method has been tested on the relatively complex groundwater system of the southern half of the Ticino basin where a steady-state groundwater flow model was set up. Obtained results show that the method allows for quantification of the amount of uncertainty in the input parameters (i.e., hydraulic conductivity and recharge). Furthermore, we show that the amount of uncertainty present in the model did not affect the conclusion that groundwater travel times in the lower Ticino basin are relatively short, while admittedly a precise estimate (i.e., with an error of less than a factor 2) might be out of reach due to the limitations introduced by the gaussian approximations of the posterior pdf of the parameters. More advanced calibration methodologies that do not require such approximations (such as Bayesian inversion by full Markov Chain Monte Carlo, see e.g. Brooks et al, 2011) might deliver more robust results, but are more computationally expensive and still dependent on the quality of the data; they will be the focus of future works. Such a method gives the modeler a better idea of the potential problems and biases that may affect in their estimates of groundwater flow-controlled variables, such as the travel time analyzed in this work, but with potential applications to solute transport processes and other conclusions drawn from groundwater flow models. Therefore, this method represents a significant step in quantifying model reliability and assessing its suitability for engineering applications in complex groundwater systems.

## Statements and Declarations